# Development of a Bio-Inspired Routing Algorithm According to Values of Solidarity and a Freirean Perspective of Engineering

Gustavo Nicolau Gonçalves, Cristiano Cordeiro Cruz and Romis Attux

***Abstract*—A routing algorithm for Señoritas Courier, a bicycle delivery cooperative in São Paulo, Brazil, composed exclusively of cis women and trans people, is presented in this paper. Unlike conventional logistics optimization, which typically focuses on cost or distance minimization, this cooperative operates under principles of solidarity, care, and equitable income distribution. The algorithm was developed through a participatory process involving cooperative members as co-designers. The classical Vehicle Routing Problem proved inadequate for this context, as it disregards individual constraints and fairness. We formulate a new variant, the Señoritas Routing Problem, which incorporates biker-specific constraints on weight, volume, and maximum distance, alongside a solidarity objective that balances route lengths. A genetic algorithm is employed as the solution method. Three fitness formulations are compared: a baseline distance-minimization operator, a constrained version, and a progressive formulation that penalizes workload imbalance. Results show that the progressive constrained formulation eliminates constraint violations and reduces the standard deviation of route lengths from 7.92 km to 0.81 km, at the cost of a moderate increase in total distance. This case demonstrates that operations research can be reoriented toward solidarity values, and that participatory methodologies are essential for aligning technical solutions with the needs of worker cooperatives.**

## I. Introduction

Contrary to common sense, optimization processes and techniques are not limited to a single perspective. We are used to thinking of optimization solely in terms of speed, cost reduction and profit maximization. However, by directing our attention to other forms of economy and alternative ways of organizing people, new optimization perspectives emerge along with new challenges and approaches: this paper tells one such story.

In recent decades, the uberization process has reshaped labor relations through digital platforms that materialize neoliberal logics, individualizing and precarious working conditions [1][2]. In Brazil, an estimated 1.5 million workers are engaged in platformized transportation and delivery services [3]. International research initiatives, such as the Fairwork project [4], have been evaluating working conditions on digital platforms across multiple countries. In its assessment of Brazilian platforms, the project examines principles of fair pay, conditions, contracts, management, and representation. The results are alarming: the highest score achieved by any platform operating in the country was only three out of a maximum of ten points, indicating systemic failures in ensuring minimum labor standards [5]. In reaction to this precarious scenario, workers have organized initiatives to create collectively owned platforms, a movement studied under perspectives such as platform cooperativism [6] and solidarity economy 2.0 [7][8].

It is within this context that the cycle delivery cooperative Señoritas Courier emerged in São Paulo, Brazil, composed exclusively of cis women and trans people. In partnership with the Technology Center of the Movimento dos Trabalhadores Sem-Teto (Homeless Workers' Movement) (MTST) and researchers from the University of Campinas (Unicamp), the cooperative began developing its own delivery platform in 2023, designed to reflect values of solidarity and care.

This paper, based on the first author's master's research under the supervision of the co-authors, presents the collaborative construction process of the routing module that integrates this platform. The developed algorithm aims to trace the daily routes of the deliverers (or bikers), as they are called in the cooperative, considering not only customer delivery requests but also the particularities of each worker: individual starting points, cargo capacities (weight and volume), and maximum distance limits.

The specificity of this context, however, posed challenges to classical routing problem modeling. The reality of Señoritas Courier required the formulation of a new variant, which we denominate the Señoritas Routing Problem (SRP), where balancing the distances traveled among bikers, expressed through the minimization of route standard deviation, assumes equivalent importance to minimizing total distance traveled.

Thus, the objectives of this paper are: (i) to present the mathematical formulation of the SRP, an extension of the classical VRP that incorporates solidarity values; (ii) to describe the solution approach developed, based on genetic algorithms; and (iii) to discuss the methodological implications of software development in solidarity economy contexts, evidencing how values can guide technical choices in operations research.

## II. Theoretical and Methodological Fundamentals

As outlined in Section I, the technology this paper analyzes – i.e., a routing algorithm developed with and for a cooperative of bicycle delivery people – can be defined as an appropriation, "customization" or widening of mainstream technology and technical disciplines – i.e. genetic algorithms, operations research and computer engineering – which embodies the cooperative's knowledge and values (such as caring, solidarity economy ideals and equity) and, in so doing, emulates and supports through its canonical use these same values. Even today, such an endeavor could sound like a corruption or "ideologization" of (pure) technology. As we will see, nonetheless, that is not at all the case, for technology is never – and can never be – (entirely) "pure" or "de-ideologized."

Evidence for this claim of "impurity" concerning technology grew mainly from the 1980s on, in the wake of Science and Technology Studies (STS) changing its focus from science to technology. Those studies have shown that technology "has politics" [9], being shaped by social forces and shaping society in turn [10][11]. This co-construction can be observed in at least two domains.

The first comprises social structures and the power arrangement they enforce. Winner's [9] well-known example of Robert Moses' low overpasses in New York City illustrates this clearly. The overpasses were built low enough to prevent buses, which were primarily used by black people in the 1930s-1950s, from accessing certain areas. Moses' racism thus shaped the overpasses, and the overpasses, in turn, enforced racial segregation. Such artifacts are therefore non-neutral, impacting society and power arrangements.

Moses' overpasses are not exceptions. We have no means of developing technology other than resorting to social values in this process. This phenomenon is called underdetermination [12][13][14], which can be defined as "any technical challenge/problem can be solved in more than one way." Underdetermination reveals that any design can take different paths, each with its own social impacts. To deal with it, technical values alone are insufficient because:

1) None of such values are absolute. They are always used within a broader framework of what is deemed acceptable by society. What was once acceptable—children in factories, unsafe boilers, polluting automobiles—is no longer [15]. The same holds for fossil fuels, medicines not tested in female bodies [16], and software that fails with minorities;

2) In most situations, no candidate solution scores higher than others in all criteria. The chosen solution reflects the criteria most relevant to those with a say in its design;

3) In some situations, technology is a disguise for social impacts. Such adoption often weakens technical values. Winner's [9] example of Cyrus McCormick's pneumatic molding machines, developed to weaken factory employees' unions, is a clear case.

In short, technology is non-neutral and underdetermined, shaping and being shaped by society. Technology must thus be politically disputed and democratized [12][13][14][15][17]. Advancing anti-racist algorithms is a possible example of democratization.

The second realm where co-construction can be found is the cultural one. That means that every culture (or, more specifically, every culture's supporting cosmology(ies)) shapes technology and, in turn, is shaped and supported by technology [18][19]. Here, what is at stake is the basis for our understanding of the underlying order of reality and our making sense of life/existence. It is culture (or its supporting cosmology(ies)) that allows us to understand how modern Americans and American Indigenous peoples come up with very different technical solutions for housing and agriculture, for instance [20][21][22]. Culture also drives Indigenous peoples from the North American West Coast, Hawaii, Australia, and New Zealand to propose protocols or guidelines for developing Artificial Intelligence (AI) solutions appropriate to those peoples [23][24][25].

The same concepts mobilized in analyzing the political/social dimension of technology – i.e., non-neutrality, underdetermination, and co-construction – are present here and play equivalent roles. Likewise, if democratizing technology is demanded in a democratic society, a multicultural world requires diversifying (on a cultural basis) technology [18]. The Indigenous protocols or guidelines for AI just mentioned are examples of such diversification.

Acknowledging technology's social and cultural dimensions – alongside its technical-scientific one, it could be said – raises the question of who and on what conditions should (or could) be incorporated into the technical design and technology development processes. Another central question is what precisely is – or should be – sought as the primary outcome of that incorporation. Let us respond to each question separately.

As for the people who should be incorporated, different emancipatory traditions such as Scandinavian Participatory Design [26] and Latin American Social Technology [27] (or Solidarity Technoscience [28]) have come to the same answer: those people affected or who will use or buy the solution being designed/developed. They should be incorporated as co-designers (as the Participatory Design tradition will call them) who bring to the designing/developing process not only their needs, urgencies, or demands but also their knowledge. Knowledge here encompasses things like their understanding of their territory, work, and/or enterprise; how they live, work, and socialize; their skills; their values and ideals; their understanding of reality and life; and their dreams concerning how life and society should be.

Incorporating co-designers this way, not only as consultants and (co-)decision-makers but also as holders of knowledge that has to be considered and incorporated into both the designing process and the final product, means that a two-way dialogue of knowledge between designers and co-designers must be in place. Through such a dialogue, designers and co-designers learn from each other, and the basis for achieving the sought socio-political democratization and/or cultural diversification is set.

Below, we will present how incorporating co-designers as holders of relevant knowledge was done to develop the routing algorithm this paper analyzes. Before that, though, it is necessary to answer the question concerning what is (or should be) the primary outcome of democratizing/diversifying technology in the (participatory) way assumed/ proposed in this paper. In short, that outcome should be to empower the people, community, or group we are working with (i.e., the Participatory Design's co-designers).

That empowerment has at least eight distinct but complementary dimensions, ranging from sociotechnical inclusion (i.e. providing access to goods or services demanded by co-designers that improve their living conditions) to political emancipation (e.g. building partnerships with different social actors in a way that leverages the co-designers' possibility of being heard and have their demands fulfilled). The more of those dimensions are addressed in the technical intervention advanced, and the more profound such intervention goes into them, the higher the empowering impacts it is likely to have [29].

The intervention that ended up with the construction of the routing algorithm analyzed in this manuscript was conceived as a case of Popular (Grassroots) Engineering (PE). PE can be defined as a way of developing social technology (primarily) through university extension, (usually) having solidarity economy ideals of self-management, care, income generation, and the like as its horizon [30]. PE heavily draws on Paulo Freire's emancipatory Popular Education [31][32] and on participatory action research [33]. PE aims at addressing empowerment's eight dimensions and at going as profoundly as possible into each one of them [29].

The work developed along with Señoritas Courier for constructing the routing algorithm was grounded in a prior investigation of software development methodologies. In particular, conventional methodologies widely adopted in the software industry were mapped, with special attention to agile approaches such as Extreme Programming and Scrum [34][35][36]. In parallel, other project experiences were examined, especially those intentionally organized so as to preserve and strengthen the knowledge, practices, and ways of life of the communities in which they were embedded [36][37][38].

As a result of this mapping effort, a methodological proposal was elaborated that combined selected agile elements with additional practices explicitly aimed at centering the development process on the cooperative's organizational logic and on the concrete needs of delivery workers [40]. Rather than assuming efficiency or performance as self-evident guiding criteria, the proposal sought to make room for organizational arrangements, working conditions, and values emerging from the cooperative itself. The resulting methodology was structured into three main phases, which are outlined in the following.

The first phase, termed Preliminary Recognition Phase, focused on understanding the cooperative's existing modes of organization and work. This phase involved establishing mutual trust among the actors involved and producing a shared understanding of current routing practices and operational constraints. Interviews and participatory workshops were conducted with some of the cooperative members in order to map workflows, decision-making processes, and the main difficulties experienced in everyday operations.

The second phase corresponded to a First Minimal Empowering Intervention. In this phase, an initial prototype of the routing algorithm was developed based on genetic algorithms, a class of evolutionary computation techniques suitable for complex optimization problems. On the one hand, the choice of genetic algorithms was motivated by their flexibility and their ability to accommodate multiple constraints beyond simple cost or time minimization. On the other hand, such heuristics are relatively accessible and can be meaningfully discussed with non-specialists, which made them particularly appropriate in a participatory development context involving the Señoritas Courier members.

The third phase, Development and Refinement, was dedicated to improving the algorithm through iterative cycles of testing and adjustment. Feedback from cooperative members played a central role in this process, guiding the tuning of parameters and the incorporation of new constraints so as to better reflect real working conditions. This iterative dynamic ensured that the solution remained aligned with the cooperative's practices and adaptable to its evolving needs.

## III. A Historical Perspective on Operations Research

The origins of operations research (OR) can be traced to the historical period of World War II. As Morse [40] documented, there was a need for systematically studying highly complex problems associated with military operations - such as search patterns for submarines and combat dynamics described by Lanchester's laws. Concurrently with these aspects, there was also a significant demand for a rigorous treatment of several problems within the broad field of economics, as exemplified in the works of Kantorovich in the Soviet Union and Koopmans and Dantzig in the United States [41]. The success of OR-based methods in these endeavors established it as a legitimate discipline for supporting rational decision making.

The advent of digital computing in the 1940s and 1950s also played a pivotal role in shaping the development of OR. The ability to perform mathematical operations with unprecedented speed enabled the construction of sophisticated models for a wide range of phenomena, allowing OR to evolve into a practical engineering discipline capable of handling a variety of large scale real world problems.

In the post-war era, operations research emerged as a cornerstone of corporate decision-making. As Morse himself anticipated, large industries promptly adopted OR methodologies to optimize efficiency. Consequently, throughout the twentieth century, the field became increasingly aligned with the demands of industrial capitalism. A recent bibliometric analysis of Operations Research, a flagship journal of the discipline, confirms this trajectory [42]. From 1952 to 2019, the most studied problems were inventory, scheduling, pricing, transportation, and risks, all directly related to corporate concerns such as logistics optimization, revenue maximization, and financial risk management. Moreover, the United States has dominated the field, publishing more than all other top ten contributing countries combined, underscoring the ties between OR and American capitalism.

The thematic evolution of OR further reveals its alignment with business interests. In the early decades, inventory and scheduling were dominant, reflecting the needs of manufacturing industries. From the 1990s onward, pricing and risk management gained prominence, mirroring the rise of financial markets. In the last decade, pricing has become the hottest topic, and new editorial areas have been created for Revenue Management and Market Analytics in response to emerging business models such as platform based markets and the gig economy [43]. This shift demonstrates that OR continues to be shaped by the questions that corporations and the military are asking, rather than by broader societal needs.

Historically, operations research has consistently sought to answer the questions posed by military and corporate actors. The Vehicle Routing Problem, for instance, was formulated to optimize delivery fleets for fuel efficiency and cost reduction, not for equitable distribution of work among drivers. While this focus has yielded powerful tools for efficiency and profit maximization, it also reveals a gap. OR has rarely been asked to address questions of fairness, care, or solidarity.

This paper argues that operations research must be developed differently, not merely applied to different problems. The demands of social movements, worker cooperatives, and solidarity economy initiatives call for new problem formulations, new solution approaches, and a different posture from researchers. The experience of the Señoritas Courier cooperative offers a concrete example of what happens when optimization is reoriented from the start toward values such as equitable distribution of income and respect for individual physical limits. It is not a rejection of OR as a discipline, but rather an invitation to begin doing it differently, guided by other questions and committed to other social actors.

## IV. Routing Problems and Genetic Algorithms

The vehicle routing problem (VRP), first formulated by Dantzig and Ramser [44], is considered a fundamental problem in combinatorial optimization. It can be considered a generalization of the Traveling Salesman Problem (TSP), hence being NP-Hard [45].

To define the VRP in mathematical terms, let us consider a graph $G = (V, E)$, in which $V$ represents a set of vertices - associated with a depot and $n$ customers' locations - and $E$ represents a set of edges - associated with a certain travel cost. There are m vehicles, each with a certain capacity $Q$, that are responsible for serving the customers, each with a certain demand. The problem is classically defined in terms of the minimization of a total cost, given that [45]: 1) each customer is visited once and only once; 2) each route begins and ends at the depot; 3) the total demand of the customers served in a route does not exceed the vehicle capacity $Q$ and 4) the length of each route does not exceed a preset limit L.

In this paper, we address a variant of the classical VRP, hereafter referred to as the Señoritas Routing Problem (SRP). The SRP emerged from a real-world collaboration with a delivery cooperative, whose daily operational logic differs significantly from the traditional VRP assumptions. Through close observation of their practices, we identified that a solution capable of truly supporting their work must account not only for efficiency but also for fairness and care among the couriers. To capture this reality, we developed an optimization problem grounded in the classical VRP framework, but extended to incorporate multiple starting points, individual capacity constraints (weight, volume and distance), and a solidarity objective that balances the distances traveled by each biker.

Formally, let a complete directed graph $G = (V, E)$ be defined, where $V = \{v_0, v_1, ..., v_n, v_{n+1}, ..., v_{n+m}\}$. Vertex $v_0$ represents the pickup point. The set $C = \{v_1, ..., v_n\}$ corresponds to the $n$ customer delivery locations, each associated with a positive demand comprising both weight $w_i$ and volume $u_i$. The set $B = \{v_{n+1}, ..., v_{n+m}\}$ represents the starting locations of the mm available bikers. Each biker $k$ has a maximum cargo capacity (weight $W_k$ and volume $U_k$) and a maximum route length $L_k$, reflecting individual physical capacities.

Each edge $(i, j) \in E$ is associated with a travel distance $d_{ij}$ (or, alternatively, a travel time or cost). The routes are constructed such that:

1. **Route Structure**: Each biker $k$ starts at their respective starting location $s_k \in B$, travels to the pickup point $v_0$ to collect the assigned items, visits a subset of customers, and returns to $s_k$.
2. **Customer Assignment**: Each customer is visited exactly once by a single biker.
3. **Capacity Constraints**: For each biker $k$, the total weight and total volume of the goods delivered do not exceed $W_k$ and $U_k$, respectively.
4. **Distance Constraints**: The total length of each route does not exceed the biker's maximum allowed distance $L_k$.

The optimization problem involves two potentially conflicting objectives:

- **Minimization of Total Distance**:

  $f_1 = \sum_{k \in B} \sum_{(i,j) \in E} d_{ij} x_{ijk}$, where $x_{ijk}$ is a binary variable indicating whether biker $k$ travels directly from location $i$ to $j$.
- **Minimization of Route Length Imbalance**:

  $f_2 = \sqrt{\frac{1}{|B|} \sum_{k \in B} (R_k - R^-)^2}$, i.e., the standard deviation of the route lengths $R_k$, where $R^-$ is the average route length among all bikers.

This multiobjective formulation ensures that the solution seeks not only operational efficiency (shorter total distance) but also solidarity among the couriers, promoting fairness by balancing the individual travel efforts.

Although the problem is formulated with two objectives, our solution approach treats it as a single-objective optimization problem by incorporating the imbalance objective as a penalty term in the fitness function. This way, the primary objective of distance minimization is prioritized while solutions with uneven route lengths are penalized, effectively enforcing fairness without requiring explicit multiobjective optimization.

In the literature, a variety of methods have been used to solve the VRP. These methods include heuristics, metaheuristics and exact methods [46]. Exact methods provide optimal solutions, whereas heuristics and metaheuristics are conceived so as to provide good quality solutions within a reasonable amount of computational effort. Among these methods, we may cite as examples methods like branch and cut, branch and bound, branch and price and dynamical programming[41].

Since the VRP belongs to the NP-Hard class, using exact methods to solve it can be excessively demanding for relatively large instances. In this case, heuristics and metaheuristic methods are necessary to allow high-quality solutions to be obtained in a feasible time. Examples of heuristic methods applicable to the VRP include genetic algorithms, particle swarm optimization, ant colony optimization, and large neighborhood search [47].

A widely adopted definition of metaheuristics is provided by Sörensen and Glover [48], who describe a metaheuristic as a high level problem independent algorithmic framework that provides a set of guidelines or strategies to develop heuristic optimization algorithms. The term is also used to refer to a problem specific implementation of a heuristic optimization algorithm according to the guidelines expressed in such a framework.

Genetic Algorithms (GAs) [46] form a class of bio-inspired metaheuristics whose modus operandi is based on the evolutionary principle of natural selection and on elements of modern genetics. These algorithms have been applied with success to a variety of combinatorial optimization problems, like the TSP and the problem focused in this work, the VRP [47].

An important feature of GAs (which is shared with other metaheuristics) is that their basic structure is the same for virtually any addressed optimization problem. This basic structure is represented in Fig. 1.

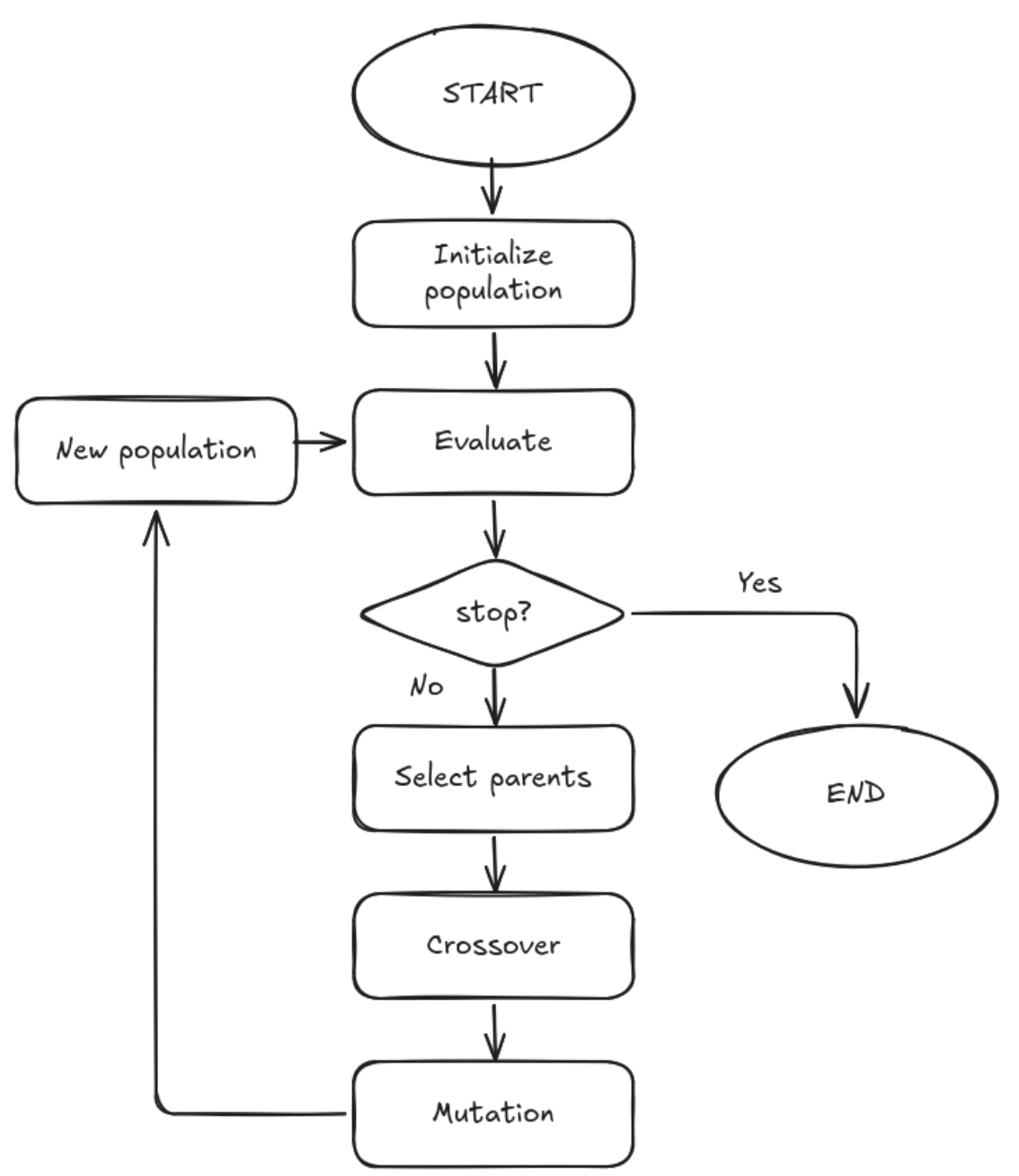

**Fig. 1**. Flowchart of a standard genetic algorithm.

In order to better understand Fig. 1, it is important to discuss some basic features of a GA [49]. The solutions to the problem being dealt with must be represented in accordance with a coding scheme. Each coded solution is thought of as a chromosome, an analogy with molecular biology. The algorithm works by considering a population of possible solutions that are subject to a cyclical process of generation of new solutions and selection.

In the context of GAs, new solutions are generated according to two main classes of operators: crossover and mutation. The crossover operator is responsible for combining (typically a pair of) existing solutions to produce new individuals (offspring). On the other hand, the mutation operator is responsible for introducing modifications of a more spurious character. Taken together, these operators allow both the refinement of the genetic material present in the population, a mechanism for local search or exploitation, and the possibility of exploring the space of possible solutions in a broader sense. These elements of local and global search must be kept in balance so as to allow the GA to have a good performance for a problem of interest.

As shown in Fig. 1, the operation of a GA begins with the initialization of the population of solutions. This initialization can be random or may incorporate available a priori knowledge [50]. After this initial step, the algorithm enters a loop of generations that comprise the evaluation of each individual, the selection of parents, the application of crossover and mutation operators, and the formation of a new population. The evaluation stage quantifies how well each solution addresses the optimization problem, assigning a fitness score that guides the subsequent selection process [49]. The selection mechanism then favors individuals with higher fitness, typically using probabilistic methods such as roulette wheel selection, ensuring that better solutions have a greater chance of propagating their characteristics to future generations. Crossover combines genetic material from selected parents to produce offspring that inherit features from both, while mutation introduces small random modifications to maintain genetic diversity and prevent premature convergence to suboptimal solutions [50]. This cycle repeats until a stopping criterion is met, such as a maximum number of generations or convergence of the population.

The use of a metaheuristic optimization algorithm is relevant in a project of this kind not only in view of the good performance of GAs in vehicle routing tasks. Their general and relatively intuitive modus operandi is also a valuable tool in allowing users who are not familiar with the mathematical foundations of optimization theory to internalize the manner whereby the problem is solved, which is an empowering tool. Unlike exact methods that require understanding of complex mathematical formulations, a GA provides a visual and conceptual framework that mirrors natural processes. This accessibility enables members of the cooperative to grasp how the routing solution is constructed, to trust its outputs, and to suggest modifications based on their practical experience. In this sense, the choice of a metaheuristic is not merely a technical decision but also a political one, as it democratizes access to the optimization process and reinforces the participatory ethos of the cooperative.

## V. Algorithm Development

The first stage of the project was dedicated to establishing the relationship between the researcher and the partner group, the bicycle delivery cooperative Señoritas Courier. This initial moment aimed not only at building trust but also at gaining an in-depth understanding of the cooperative's day-to-day functioning, particularly its organizational logic and its practices for logistical planning and route construction.

Throughout this stage, it became clear that the manual construction of routes by the bikers involved a broad and articulated set of criteria. Beyond the distances to be covered and delivery capacity, considerations included how much each biker could physically carry and pedal, the different starting points of each workday, and conditions related to both physical and mental health. Another central element in this calculation was the remuneration model adopted by the cooperative, which is based on distance traveled rather than on the number of deliveries completed. This choice reflects an effort to remunerate the work actually performed, recognizing the effort involved in longer routes and avoiding purely quantitative incentives.

This process revealed a level of complexity that appeared to exceed the routing solutions typically found in academic literature and commercial applications. At least at the outset, there was no readily available solution capable of adequately accommodating the set of criteria described by the cooperative. In this context, a genetic algorithm approach was selected due to its flexibility and its suitability for incremental development processes, in which parameters can be progressively incorporated, adjusted, or removed until a solution considered satisfactory by the cooperative members themselves is achieved.

The second stage of the project, corresponding to the First Minimal Empowering Intervention, focused on implementing an initial version of the routing algorithm after the problem had been delineated and the genetic algorithm approach selected. The objective at this stage was to construct a functional baseline solution whose components could be progressively refined in subsequent iterations.

The first design decision concerned the computational representation of candidate solutions. Chromosomes were encoded as vectors in which each position represents a delivery address and the value assigned to that position identifies the biker responsible for that delivery, as shown in Fig. 2. This encoding simplifies the crossover and mutation operations by avoiding explicit feasibility checks, thereby reducing their computational complexity. However, it does not explicitly represent the order in which deliveries are performed. To address this limitation and enable distance calculations, a heuristic was incorporated at the decoding stage to compute the shortest route for each set of deliveries assigned to a given biker. This approach mirrors the cooperative's own routing practice, in which deliveries are first allocated to bikers and routes are only defined afterward.

| 0 | 1 | 2 | 3 | 4 | 5 |
|---|---|---|---|---|---|
| 0 | 2 | 0 | 1 | 2 | 1 |

**Fig. 2**. Example of chromosome encoding for a candidate solution. Each position represents a delivery point; the value assigned to it identifies the biker responsible for that delivery.

The remaining components of the genetic algorithm were then implemented. Selection followed an elitist strategy in which the best-performing individual was always retained and combined with another individual randomly chosen from the population. For crossover, although ordered crossover was initially considered, a two-point crossover operator proved more suitable given the adopted encoding scheme [51]. Mutation was implemented as point mutation affecting a single gene. Through preliminary tests, the mutation rate was calibrated and set to 10%, as this value provided the best trade-off between global exploration and local exploitation, whereas higher rates introduced excessive randomness and impaired convergence.

The final implementation choice for this first version concerned the fitness evaluation stage shown in (1). Candidate solutions were ranked exclusively according to the total distance traveled across all delivery routes, with preference given to solutions that minimized this value. For each biker i, the total distance traveled along the route assigned to that biker ($d_i$) was computed. The fitness score of a candidate solution ($D_{max}$) was then obtained by summing the individual route distances ($d_i$) across all bikers considered in the solution, where n denotes the total number of bikers. In this formulation, each biker contributes equally to the overall score, and lower values of $D_{max}$ indicate more efficient solutions in terms of total traveled distance. This evaluation operator, termed max distance, thus directs the evolutionary process toward minimizing the cumulative distance traveled by the cooperative as a whole and establishes a baseline criterion for subsequent refinements.

$$D_{max} = \sum_{i=1}^{n} d_i \qquad (1)$$

Upon entering the final stage, Development and Refinement, the algorithm was tested using real delivery order data provided by the cooperative. This made it possible to compare the routes generated by the algorithm with those manually constructed by the cooperative in its daily operations. These experiments revealed several limitations in the responses produced by the first version of the algorithm. In particular, workload was unevenly distributed among bikers, the number of assigned deliveries often exceeded feasible limits in terms of weight and volume, and the maximum distances bikers are willing or able to cycle in a workday were frequently violated.

The analysis of these results indicated that the component with the greatest impact on solution quality was the fitness evaluation stage, as it is responsible for ranking candidate solutions and, consequently, guiding the evolutionary process. Although the initial max distance operator effectively minimized total traveled distance, it proved insufficient to capture other constraints and priorities that are central to the cooperative's routing practices.

In response to these findings, an iterative and participatory refinement process was established. New fitness operators were proposed, implemented, and tested using real-world data, and the resulting solutions were then jointly analyzed with cooperative members. Based on this collective assessment, adjustments were defined and the experimentation cycle restarted. This process was repeated until the algorithmic outputs were considered satisfactory by the cooperative. In total, two additional versions of the fitness operator were developed beyond the initial max distance formulation.

Following the max distance operator, a first revised fitness formulation was proposed in order to explicitly account for constraint violations observed in the initial experiments. This new formulation introduced penalty mechanisms for candidate solutions that infringed key problem constraints identified together with the cooperative. The selected constraints included maximum total distance per biker, maximum route distance per biker, and limits related to volume and weight of the assigned deliveries.

Importantly, these constraints were not defined as global thresholds, but as biker-specific parameters, reflecting individual differences in physical capacity, equipment, and daily conditions; for example, maximum allowable weight, volume, and route distance varied from one biker to another.

In this formulation, each violated constraint adds a penalty($p_i$) to the score of the candidate solution, increasing its fitness value and thus reducing its likelihood of being selected in subsequent generations. The overall score is computed by adding these penalties, weighted by empirically defined coefficients $\omega_i$, to the original maximum distance score $D_{max}$. The penalty weights were calibrated through an iterative process of experimentation, adjustment, and comparison with reference routing solutions manually produced by Señoritas Courier. In this process, $p_i$ assumes a value of 1 when the corresponding constraint is violated and 0 otherwise. This second fitness operator was termed constrained max distance and is formalized in (2).

$$Score = D_{max} + \sum_{i=1}^{n} \omega_i p_i \qquad (2)$$

By incorporating explicit penalties, this operator allowed the evolutionary process to progressively discard infeasible or undesirable solutions, while still preserving the distance-based optimization logic of the initial formulation. Nevertheless, important limitations remained. In particular, the operator did not adequately address the uneven distribution of work among bikers, and the binary nature of the penalty mechanism, in which constraints were either violated or not, proved insufficiently expressive to capture degrees of imbalance or partial violations. As a result, solutions that formally respected the imposed constraints could still exhibit substantial disparities in workload, motivating the development of more refined fitness formulations in subsequent iterations.

The final fitness formulation tested in the evaluation stage extended the changes introduced in the second version by incorporating a progressive penalty aimed at reducing workload imbalance among bikers. In addition to penalizing hard constraint violations, this formulation introduced a mechanism that penalizes candidate solutions proportionally to the disparity in delivery distances assigned to different bikers, thus explicitly promoting greater equity in workload distribution.

More specifically, the total delivery distance is computed for each biker in a candidate solution, and the variation among these distances is quantified by a dispersion measure denoted by α, computed as the range (or standard deviation) of route distances. A penalty weighted by a factor q, proportional to this variation, is then added to the fitness score. When delivery distances are relatively similar across bikers, the resulting penalty is negligible. Conversely, as disparities increase, the penalty grows accordingly, discouraging solutions with highly uneven workload distributions. As in the construction of the previous operator, both the variation measure and the penalty weight were determined empirically through iterative experimentation. This third fitness operator was termed progressive constrained max distance and is formalized in (3).

$$Score = D_{max} + \sum_{i=1}^{n} \omega_i p_i + \alpha q \qquad (3)$$

In summary, this formulation combines penalties for severe constraint violations that may render routes infeasible, such as excessive weight or volume, with a continuous penalty that reflects how evenly delivery distances are distributed among bikers. The empirical calibration of penalty weights and variation factors relied both on reference routing solutions provided by Señoritas Courier and on the cooperative's qualitative assessment of the routes generated by the algorithm. Parameter values were iteratively adjusted until the resulting solutions were considered satisfactory by the cooperative. The comparative performance of these fitness formulations is analyzed in the subsequent Results section.

## VI. Results

In order to track the evolution of the algorithmic solution not only qualitatively, through discussions with the cooperative, but also quantitatively, a comparative experiment was designed focusing on different versions of the fitness operator. The experiment compared three evaluation-stage formulations: max distance, constrained max distance, and progressive constrained max distance.

For each fitness operator, the genetic algorithm was executed 1000 times using the same standardized test case. This procedure allowed the computation of average values for a set of solution quality metrics, reducing the influence of stochastic effects inherent to evolutionary algorithms. The selected metrics were as follows:

1) **Average total route distance**
   The total distance traveled by all bikers, including the trajectory from their respective homes to the pickup point, the delivery routes, and the return to their homes.
2) **Average delivery route distance**
   The total distance traveled by all bikers considering only the delivery segment, from the pickup point to the final delivery location, excluding travel to and from their homes.
3) **Percentage of constraint violations**
   The proportion of executions, among the 1000 runs, that produced solutions violating one or more problem constraints, thereby compromising feasibility.
4) **Average standard deviation of delivery route distances among bikers**
   A measure of how evenly delivery distances are distributed, which serves as a proxy for workload and income equity, given that remuneration in the cooperative is based on distance traveled.

The first two metrics assess the effectiveness of each fitness formulation in minimizing traveled distances. The third metric captures the feasibility of the generated solutions, since routes that violate constraints cannot be adopted in the cooperative's daily operations. The fourth metric evaluates the degree of workload balance among bikers, a central concern for the cooperative due to its distance-based remuneration model.

The test scenario used in this experiment was provided by Señoritas Courier and consisted of a real delivery service involving two bikers, a single pickup point, and nine distinct delivery addresses. The comparative results obtained from this experiment are presented in Table I.

TABLE I
COMPARATIVE RESULTS FOR THE THREE FITNESS OPERATORS

| Fitness operator | Avg. total route (km) | Avg. delivery route (km) | Constraint violation (%) | Avg. std. dev. among bikers (km) |
|---|---|---|---|---|
| Max distance | 62.10 | 31.44 | 41.2 | 7.92 |
| Constrained max distance | 68.66 | 33.03 | 0 | 2.40 |
| Progressive constrained max distance | 72.72 | 36.32 | 0 | 0.81 |

The results show a clear trade-off between total traveled distance and constraint satisfaction. While the max distance operator yields shorter routes on average, it also presents a high rate of infeasible solutions. In contrast, the constrained and progressive constrained formulations eliminate constraint violations at the cost of increased total distance, with the latter achieving the most balanced workload distribution among bikers.

## VII. Final Remarks

The computational and mathematical techniques developed once again proved efficient for optimizing routing problems, as evidenced in this case of the Señoritas Courier cooperative. While the manual route planning process carried out by the cooperative demanded about one to two working days, the algorithmic proposal we developed performs the same processing in a matter of seconds, delivering results compatible with the quality expected by the cooperative.

However, the path to achieving this result could not follow the same trails already known in Operations Research. As presented, the cooperative rejected traditional modeling and solution approaches based on the classical VRP because they generated routes that disregarded the individual limits of the bikers and overlooked the equitable distribution of work among them. By the end of the process, we understood that the success of the proposed solution was associated with a few essential factors, which are detailed in sequence.

The first key point for the adequate development of the solution was the methodological approach adopted throughout the project. Instead of starting from a top-down or consultancy stance, which would impose ready-made solutions and expect the cooperative to adapt to them, we opted for an extensionist and dialogical approach inspired by Paulo Freire's vision, in which knowledge is built in partnership rather than imposed unilaterally.

The process was organized in three stages that sought to give protagonism to the cooperative's knowledge: a first stage of attentive listening to understand their needs and ways of operating; a second stage of proposing an initial open solution, conceived to be modified and evolved; and a third stage of participatory evolution, in which members of the cooperative gave their opinions on the results generated, validated the proposed routes, and suggested improvements. Working with real data from the cooperative and practicing listening not only in the initial phase but as a continuous movement throughout the entire development were essential factors for the final solution to genuinely engage with the group's reality and values.

From this stance and listening we arrived at the second key point: the classical VRP and its most common variations were incompatible with the group's needs, and therefore a new variant of the routing problem had to be modeled. Constructing a new problem allowed us to incorporate individual constraints of weight, volume, and maximum distance as a way to more fully respect the physical capacities of the cooperative's bikers, translating to the algorithm that different bodies with different bicycles have different needs, yet can still compose the delivery team and the final routes.

In addition to the constraints incorporated into the modeling, it was essential to add a second optimization objective in order to address the need for division of labor and income that guides the cooperative. In this work environment where financial distribution is based on distance traveled rather than on the number of deliveries, and where there is a concern to offer more equitable earning opportunities, optimizing solely for minimum total distance is not sufficient. Initial results with simpler evaluation operators showed the disparity in distances among the bikers, observed through standard deviation, which in this context also signifies differences in income. Adding a second optimization objective increased the complexity of the solution to be developed, yet it was essential for the algorithmic responses to be aligned with the cooperative's values and to be accepted by them.

These factors make evident that building technology aligned with solidarity values is not merely a technical matter, nor is it resolved by applying ready-made formulas. It is a process that demands a methodological repositioning that moves from imposition toward partnership, as well as a technical repositioning that does not shy away from finding new models or solutions for new problems. More than that, it requires taking seriously the groups with which one works, understanding that technology should serve to strengthen the values and practices that already exist, not to replace or discipline them. The experience with Señoritas Courier suggests that a good first step, perhaps the most important one, is to listen, to recognize protagonism, and to be willing to build together.

## Acknowledgment

R. Attux thanks CNPq (308425/2023-5) for the financial support.
This study was financed in part by the Coordenação de Aperfeiçoamento de Pessoal de Nível Superior (CAPES), Brazil – Finance Code 001.

**Gustavo Nicolau Gonçalves** was born in Mauá, São Paulo, Brazil, on January 6, 1996. He received the B.S. degree in computer engineering from the University of São Paulo (USP), São Carlos, Brazil, in 2019, and the M.S. degree in electrical engineering from the University of Campinas (UNICAMP), Campinas, Brazil, in 2024.

He is currently pursuing the Ph.D. degree in electrical engineering at the University of Campinas (UNICAMP), Campinas, Brazil. His research interests include evolutionary computation, vehicle routing problems, and the development of technology for solidarity economy and participatory design initiatives.

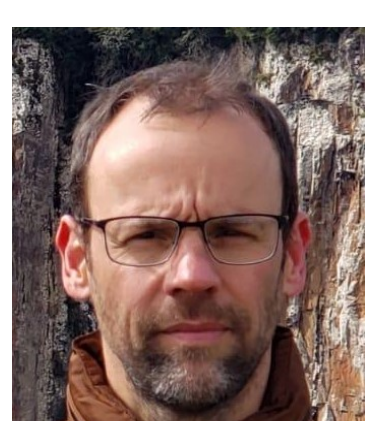

**Cristiano Cordeiro Cruz** was born in Rio de Janeiro, Brasil, on 1977. He received the B.S. and M.S. degrees in electrical engineering from the University of Campinas (UNICAMP), Campinas, Brazil, in 1999 and 2002, respectively, and the Ph.D. degree in philosophy from the University of São Paulo (USP), São Paulo, Brazil, in 2017.

He is currently a Visiting Professor with the Instituto Mauá de Tecnologia, São Caetano do Sul, Brazil. He is also a Visiting Researcher at the School of Electrical and Computer Engineering (FEEC), University of Campinas (UNICAMP). His research interests include engaged extension projects, engineering education, and the understanding of emancipatory technical practices. He is a member of the Laboratory of Citizenship and Social Technologies (LabCTS) at the Instituto Tecnológico de Aeronáutica (ITA).

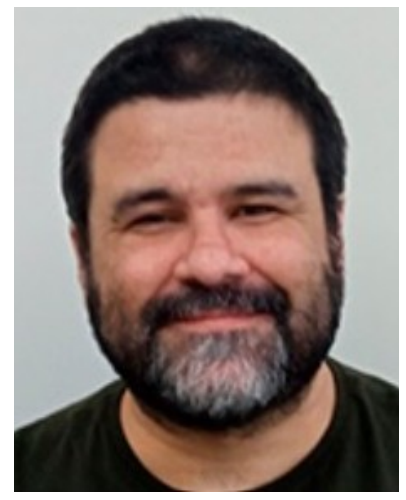

Romis Attux was born in Goiânia, Brazil, in 1978. He received the titles of Electrical Engineer (1999), Master in Electrical Engineering (2001) and Doctor in Electrical Engineering (2005) from the University of Campinas (UNICAMP).

Currently, he is an Associate Professor of the same university and a CNPq Researcher (Level 1C). His main research interests are signal processing, machine learning, brain–computer interfaces, optimization and ethical aspects of AI.